\documentclass{article}
\usepackage{spconf,amsmath,graphicx}
\usepackage{subfigure}
\usepackage{booktabs}
\usepackage{enumitem}
\setlist{nosep, leftmargin=14pt}

\usepackage{mwe} 
\usepackage{dirtytalk}
\def\x{{\mathbf x}}

\title{Multi-modal unsupervised brain image registration using edge maps}
\name{Vasiliki Sideri-Lampretsa\textsuperscript{1, 2}, Georgios Kaissis\textsuperscript{1, 2, 3}, Daniel Rueckert\textsuperscript{1, 2, 3}}
\address{\textsuperscript{1} Technical University of Munich (TUM), Munich, Germany \\
\textsuperscript{2} Klinikum Rechts der Isar, Munich, Germany \\
\textsuperscript{3} Imperial College London, UK}
\begin{document}
%
\maketitle

\begin{abstract}
\label{sec: abstract}

Diffeomorphic deformable multi-modal image registration is a challenging task which aims to bring images acquired by different modalities to the same coordinate space and at the same time to preserve the topology and the invertibility of the transformation. Recent research has focused on leveraging deep learning approaches for this task as these have been shown to achieve competitive registration accuracy while being computationally more efficient than traditional iterative registration methods. In this work, we propose a simple yet effective unsupervised deep learning-based {\em multi-modal} image registration approach that benefits from auxiliary information coming from the gradient magnitude of the image, i.e. the image edges, during the training. The intuition behind this is that image locations with a strong gradient are assumed to denote a transition of tissues, which are locations of high information value able to act as a geometry constraint. The task is similar to using segmentation maps to drive the training, but the edge maps are easier and faster to acquire and do not require annotations. We evaluate our approach in the context of registering multi-modal (T1w to T2w) magnetic resonance (MR) brain images of different subjects using three different loss functions that are said to assist multi-modal registration, showing that in all cases the auxiliary information leads to better results without compromising the runtime.

\end{abstract}
\begin{keywords}
multi-modal registration, inter-subject, gradient magnitude, deep-learning registration, unsupervised learning
\end{keywords}
\section{Introduction}
\label{sec:intro}

Image registration is an important clinical application for many medical imaging workflows. Its goal is to find corresponding anatomical or functional locations in two or more images from the same or different subjects, acquired by one or more imaging modalities \cite{Sotiras2013DeformableMI}. In clinical applications, it is often beneficial to fuse information from multiple image modalities for diagnosis or intervention guidance due to the fact that each imaging modality depicts different aspects of the underlying anatomy and pathology.

\begin{figure}[htb]

\begin{minipage}[b]{1.0\linewidth}
  \centering
  \centerline{\includegraphics[width=8.5cm]{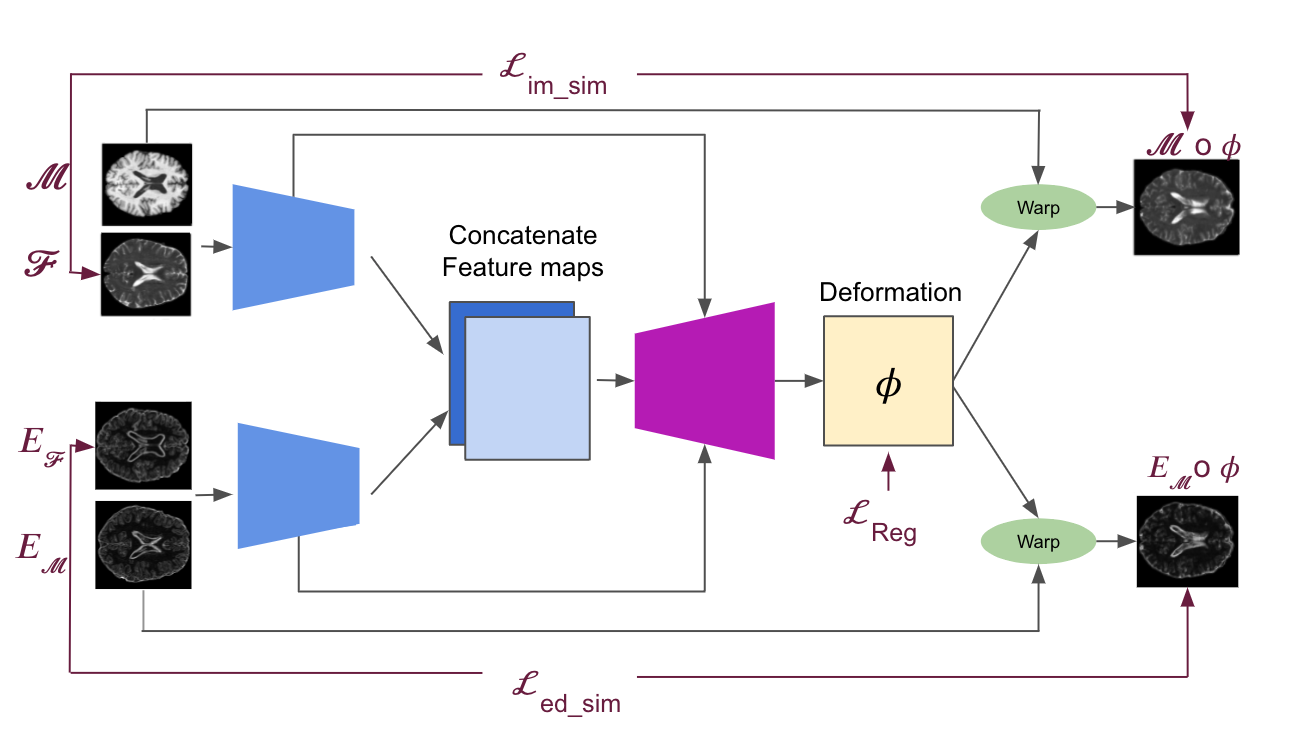}}

\end{minipage}

\caption{Network architecture. Given a moving T1w and a fixed T2w image, we first calculate their corresponding edge maps. Then, we map the fixed and moving images and their edge maps to the parameters of the transformation using a U-Net architecture which outputs the deformation field to warp the images. The loss function is a weighted combination of a loss function for image registration, a loss function for edge map registration and a regularisation loss that penalises illegal transformation configurations.}
\label{fig:larger_arch}
\end{figure}

The process of image registration involves finding the optimal geometric transformation which maximizes the correspondences across the images, i.e. given a target image $T$ and a source image $S$, image registration aims to find the optimal spatial transformation \(\phi\) which maps a location \(\x\) of \(S\) to the location with corresponding tissue or structure of \(T\) \cite{various2011Biomedical}. In traditional iterative approaches, the image alignment is achieved via a an optimisation process which normally requires many iterations to be effective. Thus, it can be time consuming. On the other hand, more recent works are exploring data-driven approaches and the incorporation of deep learning to tackle the problem \cite{Balakrishnan2019}, \cite{deVos2020}, \cite{Qin2019UnsupervisedDR}, \cite{Qiu2021LearningDA}, \cite{Mok2020LargeDD}. Although sometimes training can be time consuming, the registration step during inference is faster and thus more efficient compared to the iterative optimisation methods. 

Recent works focus on training networks in an unsupervised manner, i.e. without supervising the task with ground truth transformations. Most of these works only consider the unimodal registration task and make use of intensity based similarity metrics such us Mean Squared Error (MSE) or Cross Correlation (CC) \cite{Avants2008SymmetricDI} as image matching criterion \cite{Balakrishnan2019}, \cite{Mok2020LargeDD}, which make specific assumptions (identity or linear relationship) about the image intensities. However, these metrics are normally not suitable for the challenging problem of multi-modal image registration due to the unknown relationships between the intensity distributions and the potential differences in the geometry that each modality can capture. 

To tackle this problem, previous multi-modal approaches \cite{deVos2020}, \cite{Qiu2021LearningDA} rely on information theoretic measures such as mutual information (MI) \cite{Viola2004AlignmentBM}. However, such measures might not be suitable because they are agnostic to spatial dependency between adjacent voxels. Another approach is to reduce the multi-modal registration problem to a unimodal one \cite{Qin2019UnsupervisedDR}. Techniques like the latter are very time consuming to train and hard to control.

In this paper, we propose a novel unsupervised multi-modal image registration approach to tackle the problem of registering images of different subjects (inter-subject) acquired by two different imaging modalities (T1w, T2w MR images). We are integrating ideas from the work of Pluim et al. \cite{Pluim2000ImageRB} that combines mutual information with a term based on the magnitude and direction of the image gradients to rigidly register 3D MR, CT and PET images, the work of Qin et al. \cite{Qin2018JointLO} that uses segmentation maps as auxiliary information to guide the task of motion estimation and the work of Zhe Xu at al. \cite{Xu2021UnsupervisedMI} that performs unsupervised multimodal CT-MR image registration leveraging the deformation fields estimated from a branch aiming to align the original fixed and moving images and a branch aiming to align the corresponding gradient intensity maps.

Similarly, we propose to make use of the edge maps extracted the images, as a complementary signal, to aid the training of registration network. The main assumption is that since the anatomy depicted by the two modalities is very similar, the edge maps can serve as a geometrical constraint. At the same time, 

the edges are less \say{modality dependent}, a property that is very useful in the case of multi-modal registration due to the fact that we do not have to deal with the complicated intensity relationship between the modalities. This simple, yet, effective technique shows that the edge signal can be beneficial for multi-modal image registration.

Different from the typical one-branch unsupervised image registration network, our approach leverages the deformation field estimated from two branches that perform image and edge registration simultaneously. This is very similar to the work of Zhe Xu at al. \cite{Xu2021UnsupervisedMI}, but instead of trying to fuse the deformation fields resulting from the image and edge branches respectively, we try to estimate only one deformation field for both branches at the same time. The network architectures that we incorporated and extended are CNN networks based on Voxelmorph \cite{Balakrishnan2019} and on MIDIR \cite{Qiu2021LearningDA}. The whole pipeline of our method is depicted in \ref{fig:larger_arch}. The network consists of 2 encoders: one for encoding the features of the images and the other for encoding the features of the edge maps. Then the encoded features are passed through a decoder that outputs directly a dense stationary velocity field (SVF), in the case of Voxelmorph, or the velocities of the B-spline control points from which we can compute the transformation via B-spline tensor product and Squaring and Scaling in the case of MIDIR. 

To train both networks we employ as loss functions: local normalised cross correlation (LNCC) \cite{Avants2008SymmetricDI} as implemented in \cite{Balakrishnan2019}, normalised mutual information (NMI) \cite{Studholme1999AnOI} as implemented in \cite{Qiu2021LearningDA} and normalised gradient field (NGF) \cite{Haber2006IntensityGB}, an edge-based loss function suitable for multi-modal registration. Our experimental results prove that employing the edge maps leads to better registration accuracy without a huge increase in the memory requirements and the runtime.

\section{METHOD}
\label{sec:method}



\subsection{Edge Maps}
\label{ssec:edge_maps}

As stated above, the edge maps extracted from multi-modal images can be a useful auxiliary information to guide multi-modal image registration. Specifically, the edge map of a 2D (but easily extensible to 3D) image can be easily obtained by calculating the central differences between adjacent pixels. 



We decided to only use the gradient magnitude because it is sufficient to depict the underlying geometry and discard the information relating to the edge direction. The part of the network that handles the edge maps is more sensitive in capturing the structural dependencies between the two images, rather than the complex intensity relationships, and hence serves as a spatial constraint assisting the task of image registration. In practice, the edge maps are given as input to an encoder which adopts the same architecture as the one that handles the images.


\subsection{Network Architectures and Transformation}
\label{ssec:network_architecture}

In our effort to tackle the problem of multi-modal deformable image registration after the affine alignment, we decided to extend two different existing network architectures (\cite{Balakrishnan2019} and \cite{Qiu2021LearningDA}). The generic network architecture is depicted in Figure \ref{fig:larger_arch} and it is in general a convolutional U-Net \cite{Ronneberger2015UNetCN} architecture that consists of two separate encoders without shared weights. The first encoder takes as input the T1w and T2w images, while the second one takes the edge maps of those. Both encoders reduce the spatial dimensions in half at each layer, operating on coarser representations. The features coming from the two encoders are then concatenated and passed to a common decoder which consists of upsampling, convolutions and concatenation of skip connections and outputs a dense velocity field or the B-spline control points. The velocity field can be a direct output of the network as described by \cite{Dalca2018UnsupervisedLF}, or can be determined using the B-spline control points from which the diffeomorphic transformation can be computed via B-spline tensor product and Squaring and Scaling, as described by \cite{Qiu2021LearningDA}. All convolution layers use a kernel of size $3$ in all spatial dimensions and a stride of 2. The non-linearities are LeakyReLUs with a slope of $-0.2$.

\subsection{Loss Functions}
\label{ssec:loss_functions}

During the training, the moving image \(M\) and its corresponding edges \(E_M\) are warped using the transformation \(\phi\) via linear interpolation of the image intensities to generate the moving image \(M \circ \phi\) and moving edge image \(E_M \circ \phi\) respectively. These are then used to compute the image similarity loss \(\mathcal{L}_{\operatorname{im-sim}}(F, M \circ \phi)\) and the edge similarity loss \(\mathcal{L}_{\operatorname{ed-sim}}(E_F, E_M \circ \phi)\). The main goal is to find the optimal network parameters that minimise the loss function which is comprised by a term for image similarity, a term for edge image similarity and a term for the transformation regularisation on the velocity field. 

In order to compute the edge similarity loss, we chose to use Local Normalised Cross Correlation (LNCC) which is a local metric suitable for tasks with intensity variations close to liner \cite{Avants2008SymmetricDI}. On the other hand LNCC is also incorporated as an image similarity metric along with the differentiable version of Normalised Mutual Information \cite{Studholme1999AnOI} as introduced by \cite{Qiu2021LearningDA} and the edge-based metric Normalised Gradient Field (NGF) \cite{Haber2006IntensityGB}. The \(\mathcal{L}_{\operatorname{Reg}}\) is a regularisation on the velocity field to ensure smoothness and diffeomorphism \cite{Beg2004ComputingLD}.

The overall loss can be written as:
\begin{align}
&\mathcal{L}_1 = \lambda_1\mathcal{L}_{\operatorname{im-sim}}(F, M \circ \phi), \\
&\mathcal{L}_2 = \lambda_2\mathcal{L}_{\operatorname{ed-sim}}(E_F, E_M \circ \phi), \\
&\mathcal{L}_3 = \lambda_3 \mathcal{L}_{\operatorname{Reg}}, \\
&\mathcal{L} = \mathcal{L}_1 + \mathcal{L}_2 + \mathcal{L}_3
\end{align}
where \(\lambda_1, \lambda_2, \lambda_3 \geq 0\) are the weighting factors that are optimally tuned using a validation set, while considering the balance between registration accuracy and transformation regularity.

\section{Experiments}
\label{sec:experiments}

\subsection{Implementation Details}
\label{ssec:implementation_details}

We used PyTorch to implement  the deep learning networks and loss functions. The Adam \cite{Kingma2015AdamAM} optimiser is used with an initial learning rate \(1e^{-4}\) and the learning rate decay is set to 0.1 every 50 epochs. The traditional SVFFD method, which we use as a baseline, is implemented by the MIRTK (Medical Image Registration ToolKit) \cite{Schuh2014ConstructionOA}. All the time measurements are being performed on a workstation with an AMD Ryzen\textsuperscript{TM} Threadripper\textsuperscript{TM} 3960X CPU and NVIDIA\textsuperscript{\textregistered} Quadro RTX 8000 GPU.

\subsection{Tasks and Data}
\label{ssec:data}

We evaluate our work on the task of inter-subject brain MRI registration on T1w-T2w images. For this purpose, we decided to use the Cambridge Centre for Ageing and Neuroscience (CamCAN) dataset \cite{Shafto2014TheCC}, \cite{Taylor2017TheCC}. The dataset consist T1w and T2w MR 3D volumetric images of 310 subjects, with \(1\mbox{mm}^3\) isotropic spatial resolution. In our work we only use 2D slicse of the volumetric image cropped to the size of \(192 \times 192\). Although this work could directly be transferred in 3D, we use 2D images as a proof-of-concept, due to the large computational requirements when using 3D images. After affine alignment to a common MNI space using ANTs \cite{Avants2008ANTs}, the images are skull-stripped using ROBEX \cite{Iglesias2011Robex} and bias-field corrected using the N4 algorithm in SimpleITK \cite{Lowekamp2013TheDO}. For evaluation, we also acquired the segmentation of 138 cortical and sub-cortical structures (grouped into 5 groups) automatically using MALPEM \cite{Ledig2015RobustWS}. 

\subsection{Evaluation Metrics}
\label{ssec:evaluation_metrics}

In order to evaluate our method, we choose to both evaluate the accuracy and the regularity of the registration. The accuracy is being evaluated using the Dice score to measure the overlap between anatomical segmentations of fixed and moving images. On the other hand, the Jacobian determinant of the transformation is an index of the registration regularity, i.e. the ratio of points with \(J<0\) evaluates the amount of folding of the transformation and the magnitude of the gradient of the Jacobian (\(|\nabla_{J}|\)) determinant evaluates the spatial smoothness of the transformation.

\subsection{Comparison to state-of-the-art Registration}
\label{ssec:comparisons}

To present the effectiveness of the incorporation of the edge information in the training process we are comparing our method with three baseline methods. The first comparison is with MIRTK, an iterative method based on SVFFD transformation \cite{Schuh2014ConstructionOA}. Regarding the state-of-the-art deep learning methods, we decided to compare with Voxelmorph \cite{Balakrishnan2019} and MIDIR \cite{Qiu2021LearningDA} using the loss functions stated above. Additionally, our reference is the evaluation metrics of the initial affine registration.

\subsection{Results}
\label{ssec:results}
In Table \ref{tab:network_results} we present the quantitative evaluation of all models and loss functions on the inter-subject brain T1w-T2w registration task. The network architectures that end with ED depict the ones that were trained with the incorporation of the edge maps. We have trained the edge branch using LNCC and MSE as a loss function, second and third column of \ref{tab:network_results} respectively. In the 2D case images trained without the edge information, we found that LNCC performs most of the times better than NMI and NGF. Although cross correlation is not usually beneficial for multi-modal registration, its local implementation manages to outperform the other metrics. The reason for that may be that T1w and T2w images are locally closely related and hence the LNCC performs particularly well in our case. We do not expect this behaviour though when we have to combine features from two very different modalities e.g. MR (Magnetic Resonance) - CT (Computed Tomography). 

\begin{table}[htpb]
  \huge
  \centering
  \resizebox{\columnwidth}{!}{
  \begin{tabular}{c | c  c  c | c c c}
    \toprule
     {} & \multicolumn{3}{c|}{LNCC as edge loss} & \multicolumn{3}{c}{MSE as edge loss} \\
    \midrule
          Methods & Dice & $J<0$ & $|\nabla_{J}|$ & Dice & $J<0$ & $|\nabla_{J}|$ \\
    \midrule
        \multicolumn{1}{l|}{Affine} & $0.643\pm0.087$ & - & - & - \\
    \midrule
        \multicolumn{1}{l|}{SVFFD\textsubscript{LNCC}} & $0.749\pm0.042$ & $5.3e^{-4}$ & $0.025$ & $0.749\pm0.042$ & $5.3e^{-4}$ & $0.025$ \\
        \multicolumn{1}{l|}{VM\textsubscript{LNCC}} & $0.733\pm0.055$ & $22e^{-4}$ & $0.073$  & $0.733\pm0.055$ & $22e^{-4}$ & $0.073$   \\
        \multicolumn{1}{l|}{MIDIR\textsubscript{LNCC}} & $0.691\pm0.063$ & $13e^{-4}$ & $0.0187$ &  $0.691\pm0.063$ & $13e^{-4}$ & $0.0187$\\
        \multicolumn{1}{l|}{\textbf{VM\_ED\textsubscript{LNCC}}} & $\mathbf{0.754\pm0.043}$ & $63e^{-4}$ & $0.044$ & $\mathbf{0.748\pm0.065}$ & $37e^{-4}$ & $0.074$ \\
        \multicolumn{1}{l|}{MIDIR\_ED\textsubscript{LNCC}} & $0.732\pm0.070$ & $0.37e^{-4}$ & $0.02$ & $0.691\pm0.049$ & $0.67e^{-4}$ & $0.01$ \\
    \midrule
        \multicolumn{1}{l|}{SVFFD\textsubscript{NMI}} & $0.7393\pm0.081$ & $2.5e^{-4}$ & $0.024$ & $0.7393\pm0.081$ & $2.5e^{-4}$ & $0.024$ \\
        \multicolumn{1}{l|}{VM\textsubscript{NMI}} & $0.71\pm0.076$ & $0.3e^{-4}$ & $0.032$ &  $0.71\pm0.076$ & $0.3e^{-4}$ & $0.032$\\
        \multicolumn{1}{l|}{MIDIR\textsubscript{NMI}} & $0.699\pm0.073$ & $0.1e^{-4}$ & $0.013$ & $0.699\pm0.073$ & $0.1e^{-4}$ & $0.013$\\
        \multicolumn{1}{l|}{\textbf{VM\_ED\textsubscript{NMI}}} & $\mathbf{0.759\pm0.093}$ & $0.1e^{-4}$ & $0.033$ & $\mathbf{0.732\pm0.093}$ & $0.28e^{-4}$ & $0.06$ \\
        \multicolumn{1}{l|}{MIDIR\_ED\textsubscript{NMI}} & $0.721\pm0.0731$ & $0.5e^{-4}$ & $0.015$ & $0.69\pm0.068$ & $0.4e^{-4}$ & $0.06$ \\
    \midrule
        \multicolumn{1}{l|}{VM\textsubscript{NGF}} & $0.672\pm0.11$ & $0.16e^{-4}$ & $0.0035$ & $0.672\pm0.11$ & $0.16e^{-4}$ & $0.0035$ \\
        \multicolumn{1}{l|}{MIDIR\textsubscript{NGF}} & $0.652\pm0.07$ & $1.2e^{-4}$ & $0.0042$ & $0.652\pm0.07$ & $1.2e^{-4}$ & $0.0042$ \\
        \multicolumn{1}{l|}{\textbf{VM\_ED\textsubscript{NGF}}} & $\mathbf{0.694\pm0.087}$ & $0.1e^{-4}$ & $0.021$ & $0.645\pm0.098$ & $0.1e^{-4}$ & $0.005$\\
        \multicolumn{1}{l|}{MIDIR\_ED\textsubscript{NGF}} & $0.681\pm0.101$ & $0.23e^{-4}$ & $0.011$ & $\mathbf{0.711\pm0.064}$ & $0.04e^{-4}$ & $0.028$ \\
    \bottomrule
  \end{tabular}
  }
  \caption[Quantitative Results.]{Quantitative results for T1w-T2w inter-subject image registration. We report the average Dice score of the different anatomical structures, the Jacobian determinant of the transformation ($J<0$) whose higher number has the meaning of more folding and the Jacobian gradient magnitude ($|\nabla_{J}|$) whose lower values denote the smoother transformations.}\label{tab:network_results}
\end{table}


Moreover, we observe that Voxelmorph attains better average Dice score for almost every loss function. Nevertheless, MIDIR achieved better transformation regularity, while the traditional iterative method achieved the better Dice score outperforming the deep learning ones. The intuition behind that could be that MIRTK used SVFFD which results to a very detailed optimisation for each pair of images and a multi-resolution approach, while the deep learning methods perform one pass predictions without using multiple resolutions. On the other hand, the deep learning methods with the incorporation of the edge maps during the training managed to outperform the methods without the edges in each experiment. Interestingly, these methods also result in better Dice scores than the SVFFD method, with Voxelmorph to achieve the best results with every loss. These results confirm our hypothesis and actually strongly supports that something very simple to calculate as the gradient of the image edges can be useful as auxiliary information to drive the registration.


\begin{figure}
    \centering
    \subfigure[Moving]{\includegraphics[width=0.12\textwidth]{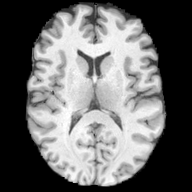}} 
    \subfigure[Fixed ]{\includegraphics[width=0.12\textwidth]{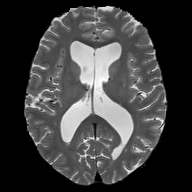}} 
    \subfigure[\(M \circ \phi\)]{\includegraphics[width=0.12\textwidth]{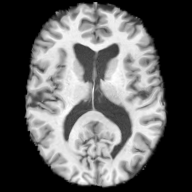}}
    \subfigure[Moving edges ]{\includegraphics[width=0.12\textwidth]{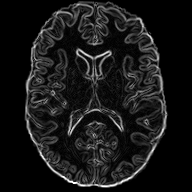}}
    \subfigure[Fixed edges]{\includegraphics[width=0.12\textwidth]{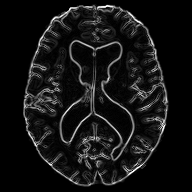}}
    \subfigure[\(E_{M} \circ \phi\)]{\includegraphics[width=0.12\textwidth]{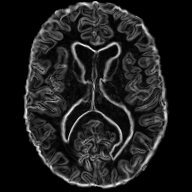}}
    \caption{T1w and T2w images with the corresponded extracted edge maps and the outcome of the deformable registration using our method.}
    \label{fig:edge_maps}
\end{figure} 

\section{Conclusion}
\label{sec:conclusion}

In this work we present a simple yet effective way to improve the results in the challenging task of the multi-modal inter-subject brain registration. We propose a registration network architecture that benefits from the usage of the magnitude of the gradient of the image, i.e. the edge maps, during its training phase. Our experimental evaluations with different loss functions and network architectures and comparing against the state of the art demonstrate that the suggested geometrical constraint can successfully assist us in our effort to improve the image registration accuracy.

\section{Compliance with ethical standards}
\label{sec:ethics}

This work uses the CamCAN dataset which was obtained in compliance with the Helsinki Declaration, and has been approved by the local ethics committee, Cambridgeshire 2 Research Ethics Committee (reference: 10/H0308/50).

\section{Acknowledgments}
\label{sec:acknowledgments}

No funding was received for conducting this study. The
authors have no relevant financial or non-financial interests to
disclose.

\bibliographystyle{IEEEbib}
\bibliography{strings,refs}

\end{document}